\def\lsim{\raise0.3ex\hbox{$<$\kern-0.75em\raise-1.1ex\hbox{$\sim$}}}
\def\gsim{\raise0.3ex\hbox{$>$\kern-0.75em\raise-1.1ex\hbox{$\sim$}}}
\def\fm{\mathrm{fm}}

\newcommand{\dbs}{}

\documentclass[12pt]{article}

\usepackage{scicite}

\usepackage{times}

\newenvironment{sciabstract}{%
\begin{quote} \bf}
{\end{quote}}

\newcounter{lastnote}

\usepackage{graphicx}
\usepackage{multirow}

\title{Ab-initio Determination of Light Hadron Masses}

\author{
S.~D\"urr$^1$, Z.~Fodor$^{1,2,3}$, J.~Frison$^4$,
C.~Hoelbling$^{2,3,4}$, \\ R.~Hoffmann$^2$,
S.~D.~Katz$^{2,3}$, S.~Krieg$^2$, T.~Kurth$^2$,\\ L.~Lellouch$^4$,
T.~Lippert$^{2,5}$, K.K.~Szabo$^2$, G.~Vulvert$^4$ \\
\\
\normalsize{$^1$NIC, DESY Zeuthen, D-15738 Zeuthen and FZ J\"ulich, D-52425 J\"ulich, 
Germany.}\\
\normalsize{$^2$Bergische Universit\"at Wuppertal, Gaussstr.\,20, D-42119 Wuppertal, 
Germany.}\\
\normalsize{$^3$Institute for Theoretical Physics, E\"otv\"os University, H-1117 Budapest, 
Hungary.}\\
\normalsize{$^4$Centre de Physique Th\'eorique\footnote{CPT 
is ``UMR 6207 du CNRS et des universit\'es d'Aix-Marseille I,
  d'Aix-Marseille II et du Sud Toulon-Var, affili\'ee \`a la FRUMAM''.
}, Case 907, Campus de Luminy, F-13288 Marseille Cedex 9, 
France.}\\
\normalsize{$^5$J\"ulich Supercomputing Centre, FZ J\"ulich, D-52425 J\"ulich, Germany.}
\\
\\
Budapest-Marseille-Wuppertal Collaboration
}

\date{}

\begin{document}
\dbs
\maketitle
\begin{sciabstract}
More than 99\% of the mass of the visible universe 
is made up of protons and
neutrons. Both particles are much heavier than their quark and gluon
constituents, and the Standard Model of particle physics should
explain this difference.
We present a full ab-initio calculation of the masses of protons,
neutrons and other light hadrons, using 
lattice quantum chromodynamics.
Pion masses down to 190 mega electronvolts are used to
extrapolate to the physical point with lattice sizes of approximately four
times the inverse pion mass. Three lattice spacings are used for a
continuum extrapolation. Our results completely agree
with experimental observations and represent a quantitative
confirmation of this aspect of the Standard Model
with fully controlled uncertainties.
\end{sciabstract}

The Standard Model of particle physics predicts a cosmological,
 quantum chromodynamics (QCD)--related 
smooth transition between a high-temparature phase dominated by quarks
and gluons and a low-temperature phase dominated by hadrons.
The very large energy densities at the
 high temperatures of the early universe have essentially
 disappeared through expansion and cooling.
 Nevertheless, a fraction of
 this energy is carried today by quarks and gluons, which are confined into 
protons and neutrons. According to the mass-energy equivalence, $E=m\cdot c^2$, 
we experience this energy as mass.
 Because more than 99\% of the mass
 of ordinary matter comes from  protons and neutrons,
 and in turn 
 about 95\% of their mass comes from this confined 
energy,
it is
 of fundamental interest to perform a controlled, ab initio calculation based on QCD to determined the hadron masses. 

QCD is a generalized version of quantum electrodynamics (QED) which describes
the electromagnetic interactions. The
Euclidean Lagrangian with gauge coupling $g$ and a quark mass of
$m$ can be written as ${\cal
  L}$=$-1/(2g^2)$Tr$F_{\mu\nu}$$F_{\mu\nu} + {\bar
  \psi}$$[\gamma_\mu (\partial_\mu + A_\mu) + m ] \psi$, where
$F_{\mu\nu}$=$\partial_\mu$$A_\nu - \partial_\nu$$A_\mu +$ [$A_\mu$,$A_\nu$].
In electrodynamics, the gauge potential $A_\mu$ is a real 
valued field, whereas in QCD it is a 3$\times$3
matrix field. Consequently, the commutator in $F_{\mu\nu}$ vanishes in QED,
but not in QCD.  The $\psi$ fields also have an additional ``color''
index in QCD, which runs from 1 to 3.  Different ``flavors'' of quarks are
represented by independent fermionic fields, with possibly different
masses.  
In the work presented here, a full 
calculation of the light hadron spectrum in QCD, 
only three input parameters are required: the light and 
strange quark masses and the coupling $g$.

The action $S$ of QCD is defined as the four-volume integral of
${\cal L}$.  Green's functions are averages of products of fields over
all field configurations, weighted by the Boltzmann factor $\exp(-S)$.
A remarkable feature of QCD is asymptotic freedom, which means 
that for high energies (that is, for energies at least 10 to 100 times higher
than that of a proton at rest)
the interaction gets weaker and 
weaker~\cite{Gross:1973id, Politzer:1973fx}, enabling perturbative 
calculations based on a small coupling parameter. 
Much less is known about the other side, where the coupling gets 
large, and the physics describing the interactions 
becomes nonperturbative.
To explore the predictions of QCD in this nonperturbative regime, the
most systematic approach is to discretize~\cite{Wilson:1974sk} 
the above Lagrangian on a
hypercubic space-time lattice with spacing $a$, to
evaluate its Green's functions numerically and to extrapolate the
resulting observables to the continuum ($a\rightarrow0$).  A
convenient way to carry out this discretization is to place the
fermionic variables on the sites of the lattice, whereas the gauge
fields are treated as $3\times 3$ matrices connecting these sites. In
this sense, lattice QCD is a classical four-dimensional statistical
physics system.

Calculations have been performed using the quenched
approximation, which assumes that the fermion determinant
(obtained after integrating over the $\psi$ fields)
is independent of the gauge field.  Although this approach omits 
the  most computationally demanding part of a full
QCD calculation, a thorough determination of the quenched spectrum
 took almost 20 years. It was shown~\cite{Aoki:1999yr} 
that the quenched theory agreed with the
experimental spectrum to approximately 10\% for typical hadron masses and
demonstrated that systematic differences were observed between
quenched and two flavor QCD beyond that level of
precision~\cite{Aoki:1999yr, Aoki:2002fd}. 

Including the effects of the light sea quarks has dramatically
improved the agreement between experiment and lattice QCD results.
Five years ago, a collaboration of
collaborations~\cite{Davies:2003ik} produced results for many physical
quantities that agreed well with experimental results. Thanks to continuous
progress since then, lattice QCD calculations can now be performed
with light sea quarks whose masses are very close to their physical
values~\cite{Aoki:2008sm} (though in quite small volumes). Other
calculations, which include these sea-quark effects in the light
hadron spectrum, have also appeared in the literature~\cite{Bernard:2001av,Aubin:2004wf,Ukita:2007cu,Gockeler:2007rx,Antonio:2006px,
  WalkerLoud:2008bp, DelDebbio:2006cn, Alexandrou:2008tn,
  Noaki:2007es}.  However, all of these studies have neglected one or
more of the ingredients required for a full and controlled
calculation. The five most important of those are, in the order that
they
will be addressed below:\\
{I.} The inclusion of the up ($u$), down ($d$) and strange ($s$)
quarks in the fermion determinant with an exact algorithm and with an
action whose universality class is
  QCD. For the light hadron spectrum, the effects
  of the heavier charm, bottom and top quarks are included in the coupling 
  constant and light quark masses.\\
{II. } A complete determination of the masses of the light ground-state, 
flavor nonsinglet mesons and octet and decuplet baryons. 
Three of these are used to fix
  the masses of the isospin averaged light ($m_{ud}$) and 
  strange ($m_s$) quark masses and
  the overall scale in physical units.\\
{III.} Large volumes to guarantee small finite-size effects and at
  least one data point at a significantly larger volume to confirm the
  smallness of these effects. In large volumes, finite-size
  corrections to the spectrum are exponentially 
  small~\cite{Luscher:1985dn,Luscher:1986pf}. As a 
  conservative rule of
  thumb $M_\pi L\gsim 4$, with $M_\pi$ the pion mass and $L$ the lattice 
  size, guarantees that
  finite-volume errors in the spectrum are around or below the percent
  level~\cite{SOM}. Resonances require special care. 
Their finite volume behavior 
  is more involved. The literature provides 
  a conceptually satisfactory framework for 
  these effects~\cite{Luscher:1990ux,Luscher:1991cf} which should be 
  included in the analysis.\\
{IV.} Controlled interpolations and extrapolations of the results
  to physical $m_{ud}$ and $m_s$ (or eventually directly simulating 
  at these mass values). Although 
  interpolations to physical $m_s$, corresponding to
  $M_K{\simeq}$495~MeV, are straightforward, 
the extrapolations to the physical value of 
$m_{ud}$, corresponding to
  $M_\pi$$\simeq$135~MeV, are difficult. They need 
  computationally intensive calculations with $M_\pi$ reaching 
  down to 200~MeV or less.\\
{V.} Controlled extrapolations to the continuum limit, requiring
  that the calculations be performed at no less than three values of the
  lattice spacing, 
in order to guarantee that the scaling region is reached. 

Our analysis  includes all five ingredients
listed above, thus providing a calculation of the light hadron 
spectrum with fully controlled systematics as follows.

{\it I.}  Owing to the key
statement from renormalization group theory that higher-dimension,
local operators in the action are irrelevant in the continuum limit,
there is, in principle, an unlimited freedom in choosing a lattice
action. There is no consensus regarding which action would
offer the most cost-effective approach to the continuum limit and to
physical $m_{ud}$. 
We use an action that improves both the gauge
and fermionic sectors and heavily suppresses nonphysical,
ultraviolet modes~\cite{SOM}.
We perform a series of 2+1 flavor
calculations: that is, we include degenerate $u$
and $d$ sea quarks and an additional $s$ sea quark. We fix $m_s$ 
to its approximate physical value. To
interpolate to the physical value, four of our simulations were
repeated with a slightly different $m_s$. We vary $m_{ud}$
in a range that extends down to $M_\pi\approx$190~MeV.

\begin{figure}[h!]
\centerline{\includegraphics*[width=12cm]{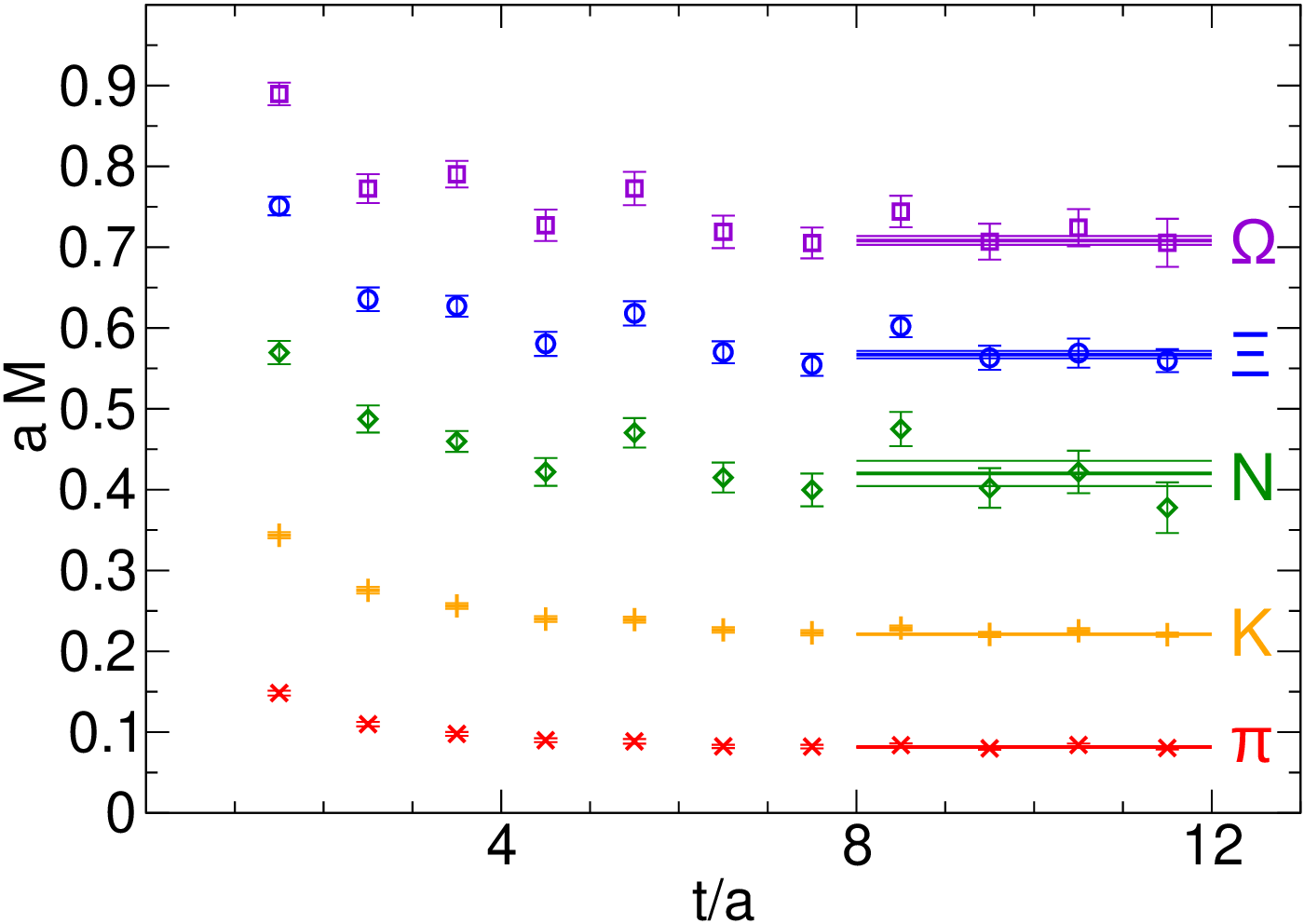}}
\caption{\label{massplat} 
\dbs
Effective masses  
$aM$=$\log[C(t/a)/C(t/a+1)]$, where $C(t/a)$ is the
correlator at time $t$, for $\pi$, $K$, $N$, $\Xi$
and $\Omega$ at our lightest simulation point
with $M_\pi$$\approx$190~MeV ($a \approx 0.085$~fm with physical
strage quark mass).
For every 10th trajectory, the 
hadron correlators were computed with 
Gaussian sources and sinks whose radii
are approximately 0.32~fm. The data points represent mean $\pm$ SEM.
The horizontal lines indicate the masses $\pm$ SEM obtained by performing
single mass correlated 
cosh/sinh fits to the individual hadron correlators with a method similar 
to that of~\cite{Bernard:2002pc}.
}

\end{figure}

{\it II.} QCD does not
predict hadron masses in physical units: only dimensionless
combinations (such as mass ratios) can be calculated.  To set the overall
physical scale, any dimensionful observable can be used.  However,
practical issues influence this choice. First of all, it should be a
quantity that can be calculated precisely and whose experimental
value is well known. 
Second, it should have a weak dependence 
on $m_{ud}$ so that its chiral behavior does not interfere
with that of other observables. Because we
are considering spectral quantities here, these two conditions should
guide our choice of the particle whose mass will set the 
scale. Furthermore, the particle should not decay under the strong
interaction. On the one hand, the larger the strange
content of the particle, the more precise the mass determination and
the weaker the dependence on $m_{ud}$. These facts 
support the use of the $\Omega$ baryon, the particle with the highest
strange content. On the other hand, the
determination of baryon decuplet masses is usually less precise than
those of the octet. This observation would suggest that the $\Xi$
baryon is appropriate.  Because both the $\Omega$ and $\Xi$ are
reasonable choices, we carry out two analyses, one with $M_\Omega$
($\Omega$ set) and one with $M_\Xi$ ($\Xi$ set). We find that for 
all three gauge couplings, $6/g^2{=}3.3$, 3.57
and 3.7, both quantities give consistent results, namely: $a{\approx}0.125$,
0.085 and 0.065~fm, respectively.  
To fix the bare quark masses, we use the mass ratio
pairs $M_\pi/M_\Omega$,$M_K/M_\Omega$ or
$M_\pi/M_\Xi$,$M_K/M_\Xi$.  We determine the masses of the baryon
octet ($N$, $\Sigma$, $\Lambda$, $\Xi$) and decuplet ($\Delta$,
$\Sigma^*$, $\Xi^*$, $\Omega$) and those members of the light
pseudoscalar ($\pi$, $K$) and vector meson ($\rho$, $K^*$) octets
that do not require the calculation of disconnected propagators. 
Typical effective masses are shown in Figure~\ref{massplat}. 

{\it III.}  Shifts in hadron masses due to the finite size of the
lattice are systematic effects. There are two different effects and we
took both of them into account.  The first type of volume dependence
is related to virtual pion exchange between the different copies of
our periodic system and it decreases exponentially with $M_\pi L$.
Using $M_\pi L{\gsim} 4$ results in masses which coincide, for all
practical purposes, with the infinite volume results [see results, for example,
for pions~\cite{Colangelo:2003hf} and for
baryons~\cite{AliKhan:2003cu,Orth:2005kq}). Nevertheless, for one of
our simulation points we used several volumes and determined the
volume dependence which was included as a (negligible) correction at
all points~\cite{SOM}.  The second type of volume dependence exists
only for resonances. The coupling between the resonance state and its
decay products leads to a non-trivial level structure in finite
volume. Based on \cite{Luscher:1990ux,Luscher:1991cf}, we
calculated the corrections necessary to reconstruct the resonance
masses from the finite volume ground-state energy and included them in
the analysis~\cite{SOM}.

{\it IV.}
Though important algorithmic developments have taken place recently
[for example \cite{Clark:2006wq,Wilcox:2007ei} and for our
setup \cite{Durr:2008rw}],
simulating directly at physical $m_{ud}$ in large enough volumes,
which would be an obvious choice, is still
extremely challenging numerically. Thus, the standard strategy
consists of performing calculations at a number of larger $m_{ud}$ 
and extrapolating the results to the physical point. To that end we 
use chiral perturbation theory and/or a Taylor expansion 
around any of our mass points~\cite{SOM}.

\begin{figure}[h!]
(A)\hspace*{8cm}(B)\\
\\
\centerline{
\includegraphics*[width=8cm]{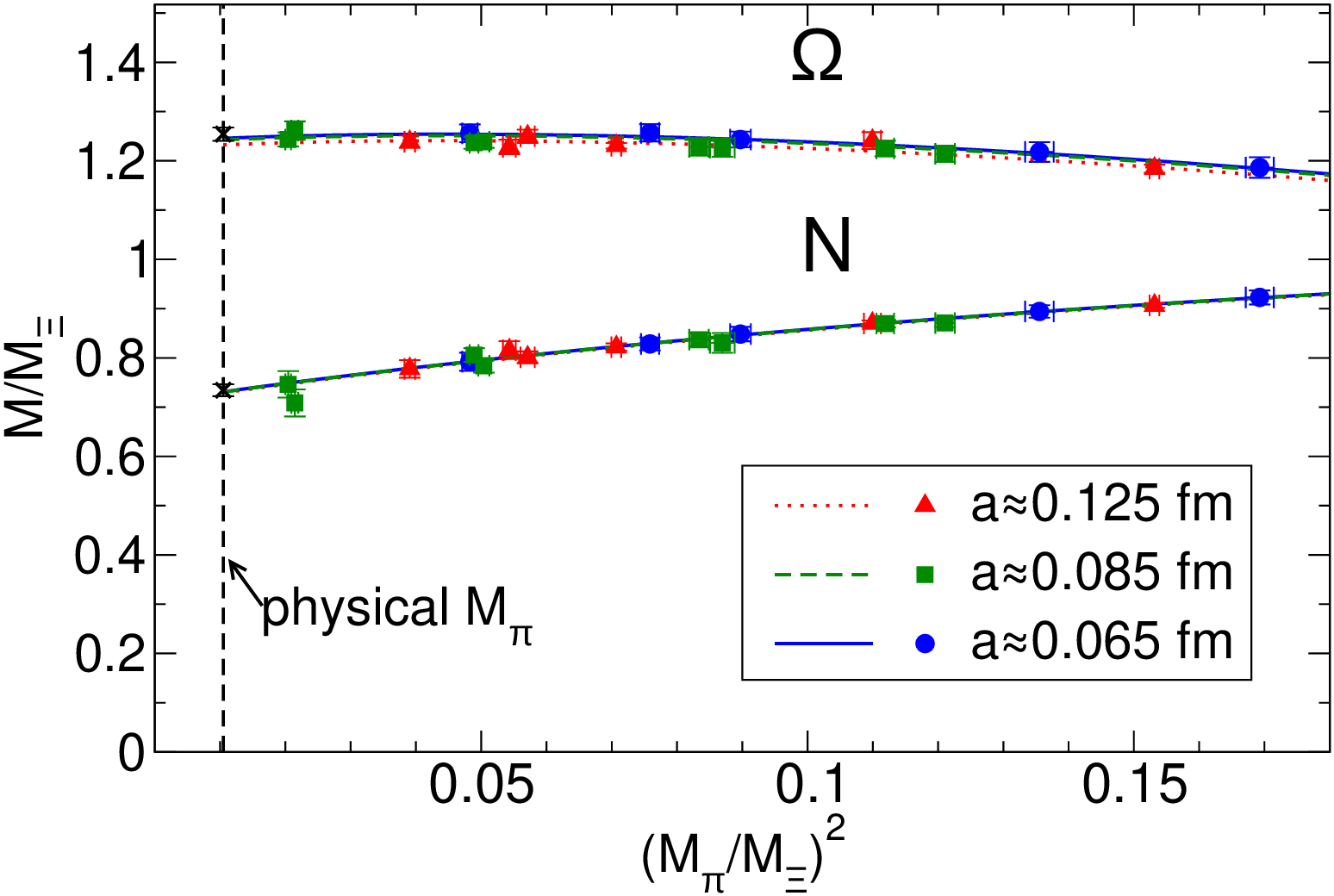}
\hspace*{0.1cm}
\includegraphics*[width=8.15cm]{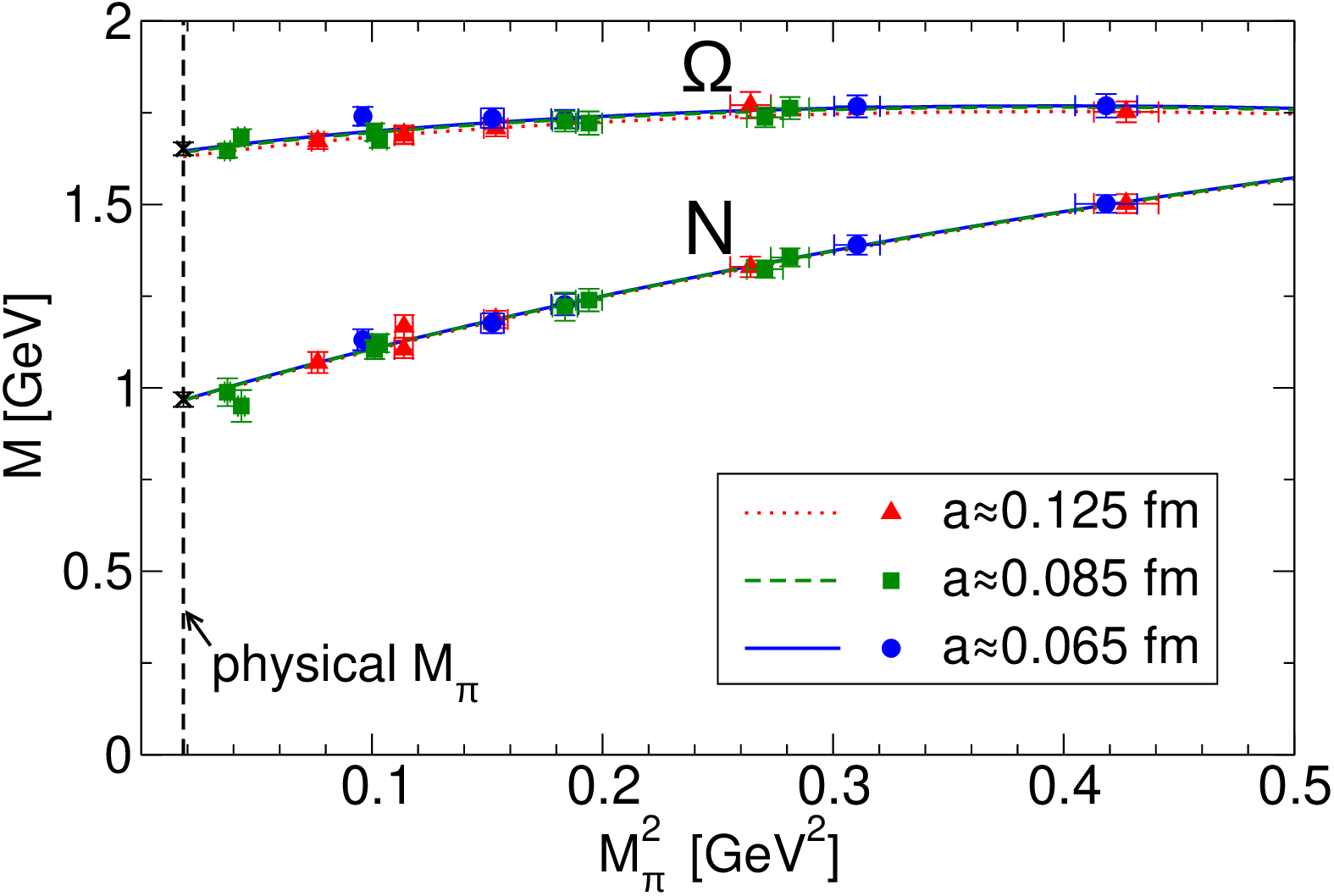}}
\caption{\label{scaling}
\dbs
 Pion mass dependence of the nucleon ($N$) and $\Omega$ 
 for all three values of the lattice spacing. (A):
 masses normalized by $M_\Xi$, 
evaluated at the corresponding simulation points.
(B): masses 
 in physical units. The scale in this case is set by $M_\Xi$ at the physical
 point.
  Triangles on dotted lines correspond to $a$$\approx$0.125~fm, 
  squares on dashed lines to
  $a$$\approx$0.085~fm and circles on solid lines to $a$$\approx$0.065~fm. 
  The points
  were obtained by interpolating the lattice results to the physical
  $m_s$ (defined by setting 2$M_K^2$-$M_\pi^2$ to its
  physical value). 
  The curves are the corresponding fits. The crosses are the continuum
extrapolated values in the physical pion mass limit.
The lattice-spacing dependence of the 
results is barely significant statistically despite the factor of 3.7
separating the squares of the largest ($a{\approx}0.125$~fm) and smallest 
($a{\approx}0.065$~fm) lattice spacings. 
The $\chi^2$/degrees of freedom values of the fits in (A) are 9.46/14 ($\Omega$) 
and 7.10/14 ($N$), whereas those 
of the fits in (B) are 10.6/14 ($\Omega$) and 9.33/14 ($N$). 
All data points represent mean $\pm$ SEM.
}
\end{figure}

{\it V.}  Our three-flavor scaling  
study~\cite{Durr:2008rw} showed that hadron masses deviate from their 
continuum values by less than approximately 1\% for lattice spacings
up to $a{\approx}0.125$~fm. Because the statistical errors of the hadron masses
calculated in the present paper are similar in size, we do not expect 
significant scaling violations here. This is confirmed by
Figure~\ref{scaling}.  
Nevertheless, we quantified and removed possible discretization 
errors by a combined analysis using results obtained at three lattice 
spacings~\cite{SOM}.

\begin{figure}[h!]
\centerline{\includegraphics*[width=12cm]{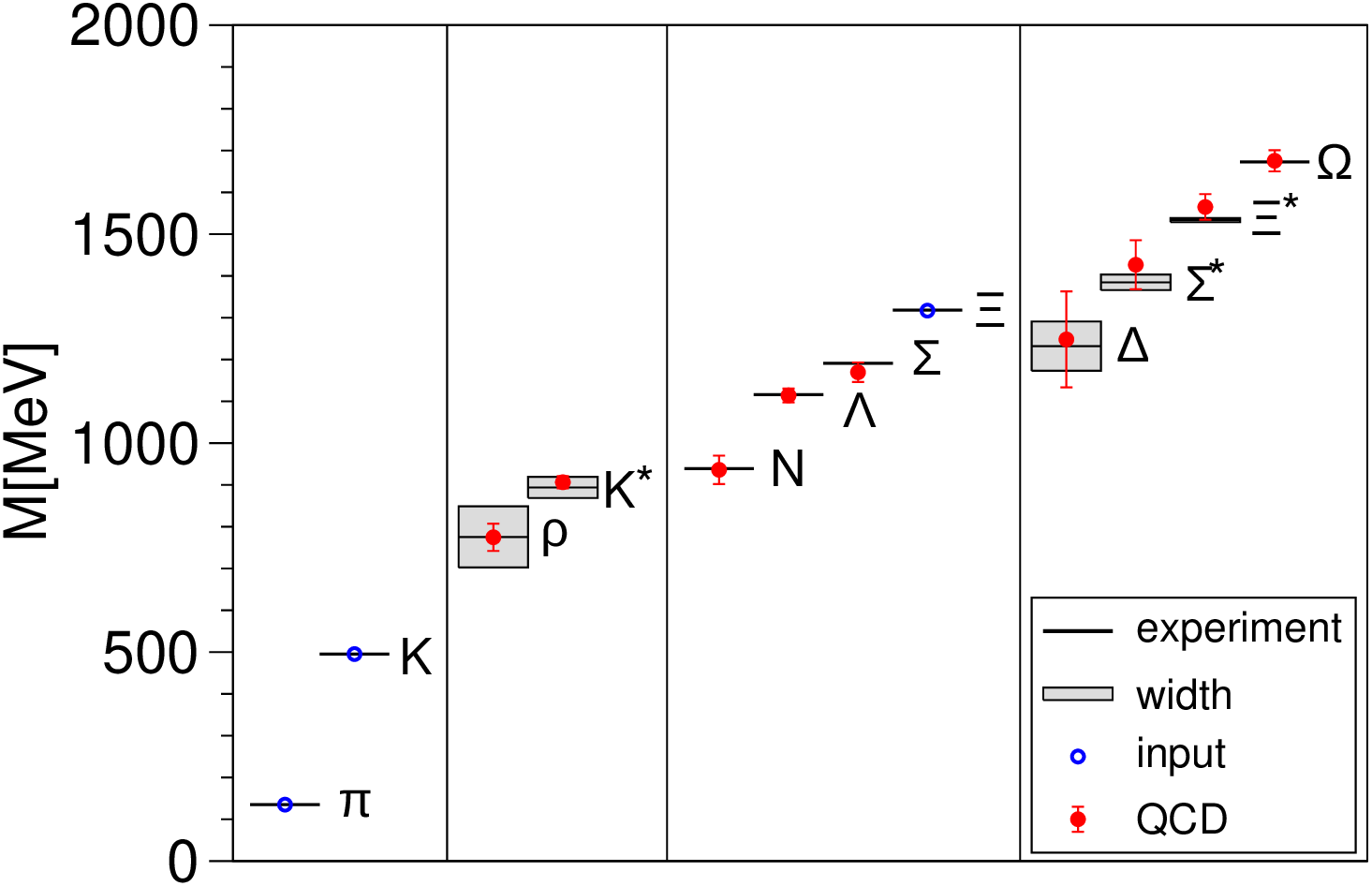}}
\caption{\label{masses}
\dbs
The light hadron spectrum of QCD. 
Horizontal lines and bands are the experimental values with their 
decay widths. Our results are shown by solid circles. 
Vertical error bars 
represent our combined statistical (SEM)
and systematic error estimates. 
$\pi$, $K$ and $\Xi$ 
have no error bars, because they are used to set the light quark mass, the  
strange quark mass and the overall scale, respectively. 
}
\end{figure}

We performed two separate
analyses, setting the scale with $M_\Xi$ and $M_\Omega$.  
The results of these two
sets are summarized in Table~\ref{results}. 
The $\Xi$ set is shown in Figure~\ref{masses}. 
With both scale-setting procedures 
we find that the masses agree with the 
hadron spectrum observed in nature~\cite{Yao:2006px}. 

Thus, our study strongly suggests that QCD is the theory of the strong
interaction, at low energies as well, and furthermore that lattice
studies have reached the stage where all systematic errors can be
fully controlled. This will prove important in the forthcoming era in
which lattice calculations will play a vital role in unraveling
possible new physics from processes which are interlaced with QCD
effects.

\clearpage

\begin{table}[h!]
\begin{center}
\begin{tabular}{llll}
\hline\hline
$X$ & Exp.\cite{Yao:2006px} & $M_X$ ($\Xi$ set) & $M_X$ ($\Omega$ set) \\
\hline
    $\rho$ & 0.775 & $0.775(29)(13)$ & $0.778(30)(33)$\\
     $K^*$ & 0.894 & $0.906(14)(4)$ & $0.907(15)(8)$\\
       $N$ & 0.939 & $0.936(25)(22)$ & $0.953(29)(19)$\\
 $\Lambda$ & 1.116 & $1.114(15)(5)$ & $1.103(23)(10)$\\
  $\Sigma$ & 1.191 & $1.169(18)(15)$ & $1.157(25)(15)$\\
     $\Xi$ & 1.318 & $1.318$ & $1.317(16)(13)$\\
  $\Delta$ & 1.232 & $1.248(97)(61)$ & $1.234(82)(81)$\\
$\Sigma^*$ & 1.385 & $1.427(46)(35)$ & $1.404(38)(27)$\\
   $\Xi^*$ & 1.533 & $1.565(26)(15)$ & $1.561(15)(15)$\\
  $\Omega$ & 1.672 & $1.676(20)(15)$ & $1.672$\\
\hline\hline
\end{tabular}
\end{center}
\caption{\label{results} 
\dbs
Spectrum results in giga electronvolts. The statistical 
(SEM) and systematic uncertainties
on the last digits are given in the first and second set of parentheses,
respectively. Experimental masses are isospin-averaged~\cite{SOM}. 
For each of the
isospin multiplets considered, this average is within at most 3.5 MeV of
the masses of all of its members.
As expected the octet masses are more 
accurate than the decuplet masses, 
and the
larger the strange content the more precise is the result. 
As a consequence the $\Delta$ mass determination is the least precise.
}
\end{table}

\vspace*{2cm}

\vspace*{2cm}

\clearpage

\centerline{{\LARGE Supplementary Online Material}}

\vspace*{1cm}

\renewcommand\citeform[1]{S#1} 
\renewcommand{\thefigure}{S\arabic{figure}}
\setcounter{figure}{0}
\renewcommand{\thetable}{S\arabic{table}}
\setcounter{table}{0}

\section*{Details of the simulations}

We use a tree-level,
$O(a^2)$-improved Symanzik gauge action\cite{S_Luscher:1985zq} and work
with tree-level, clover-improved Wilson fermions, coupled to links
which have undergone six levels of stout link 
averaging\cite{S_Morningstar:2003gk}. (The precise form of the action is presented in
\cite{S_Durr:2008rw}.)

Simulation parameters, lattice sizes and trajectory lengths after
thermalization are summarized in Table \ref{parameters}.  Note, that
we work on spatial volumes as large as $L^3{\simeq} (4\,\fm)^3$ and
temporal extents up to $T{\simeq} 8\,\fm$. Besides significantly
reducing finite-volume corrections, this choice has a similar effect
on the statistical uncertainties of the results as increasing the
number of trajectories at fixed volume. For a given pion mass, this
increase is proportional to the ratio of volumes. Thus, for $T\propto
L$, 1,300 trajectories at $M_\pi L{=}4$ are approximately equivalent
to 4,000 trajectories at $M_\pi L$=3.
(A factor $L^3$ comes from the summation over the spatial volume
required to project the hadron correlation functions onto the
zero-momentum sector and an additional $L$ comes from the fact that
more timeslices are available for extracting the corresponding hadron
mass.)

\begin{figure}
\centerline{
\includegraphics[width=8cm]{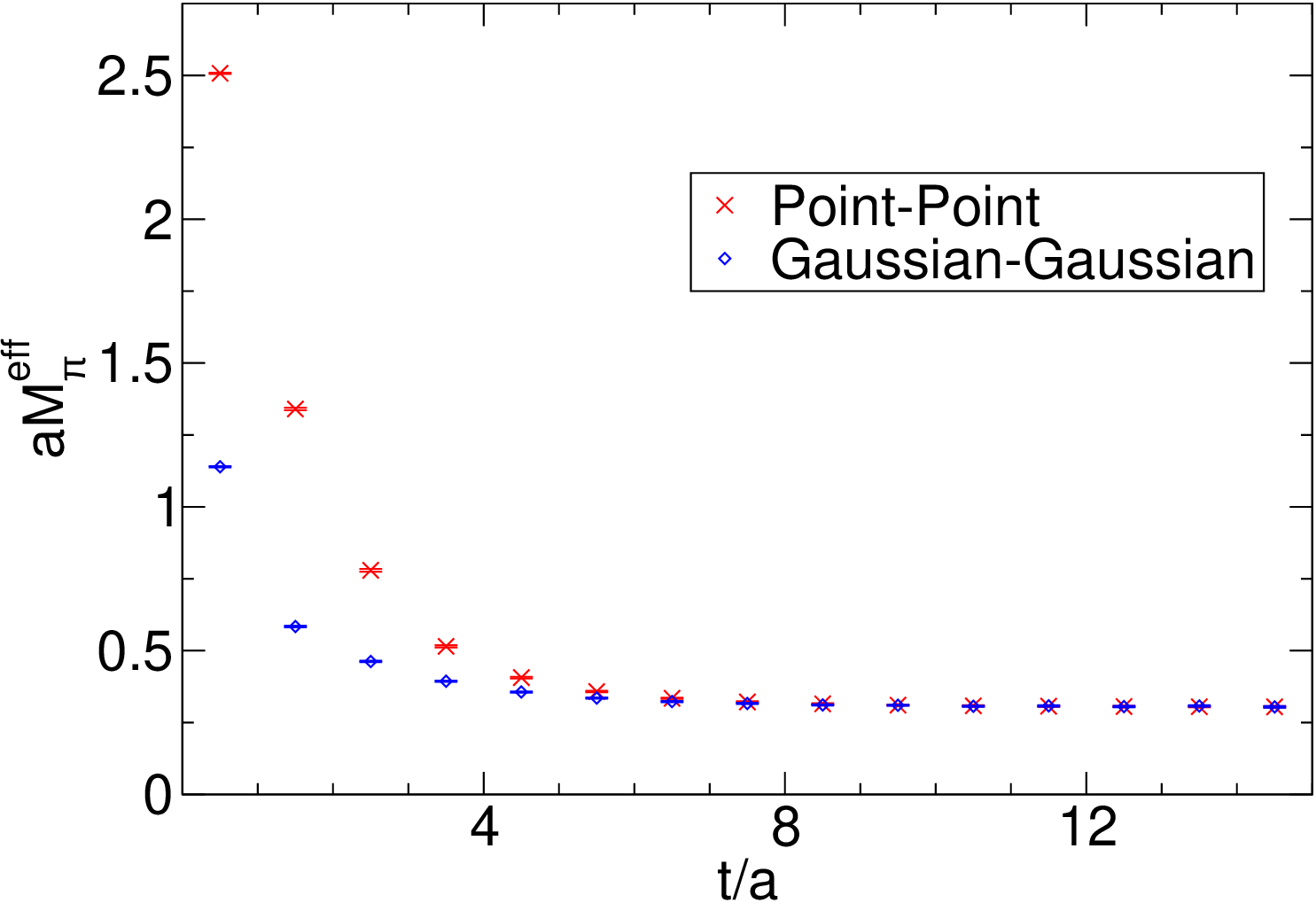}
\hspace*{0.1cm}
\includegraphics[width=8cm]{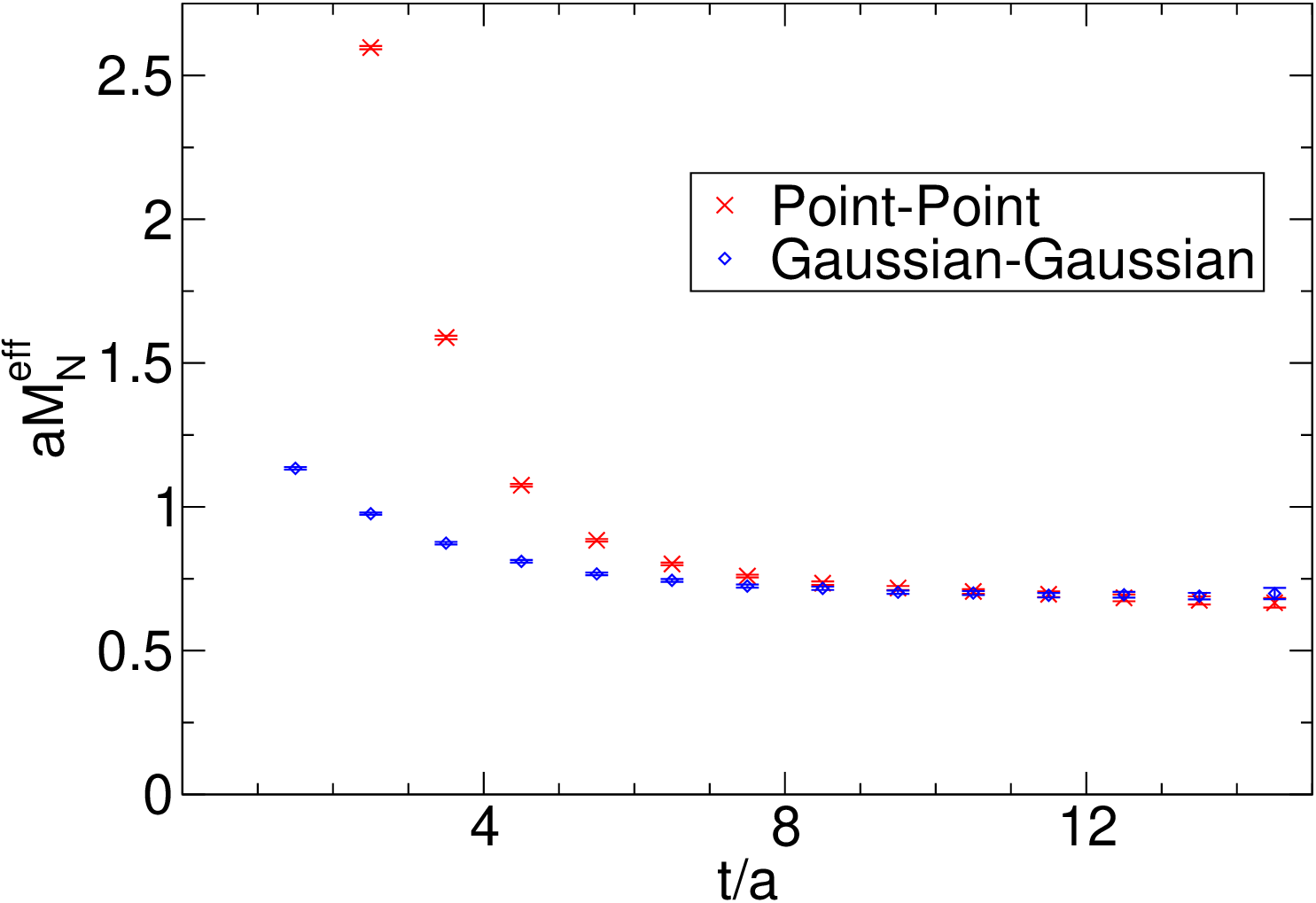}
}
\caption{\label{sources} 
\dbs
Effective masses for different source types in the pion (left panel) and
nucleon (right panel) channels. Point sources have vanishing
extents, whereas Gaussian sources, used on Coulomb gauge fixed
configurations have radii of approximately 0.32~fm. Clearly, 
the extended sources/sinks result in much smaller excited state
contamination.
}
\end{figure}

\begin{figure}
\centerline{\includegraphics[width=10cm]{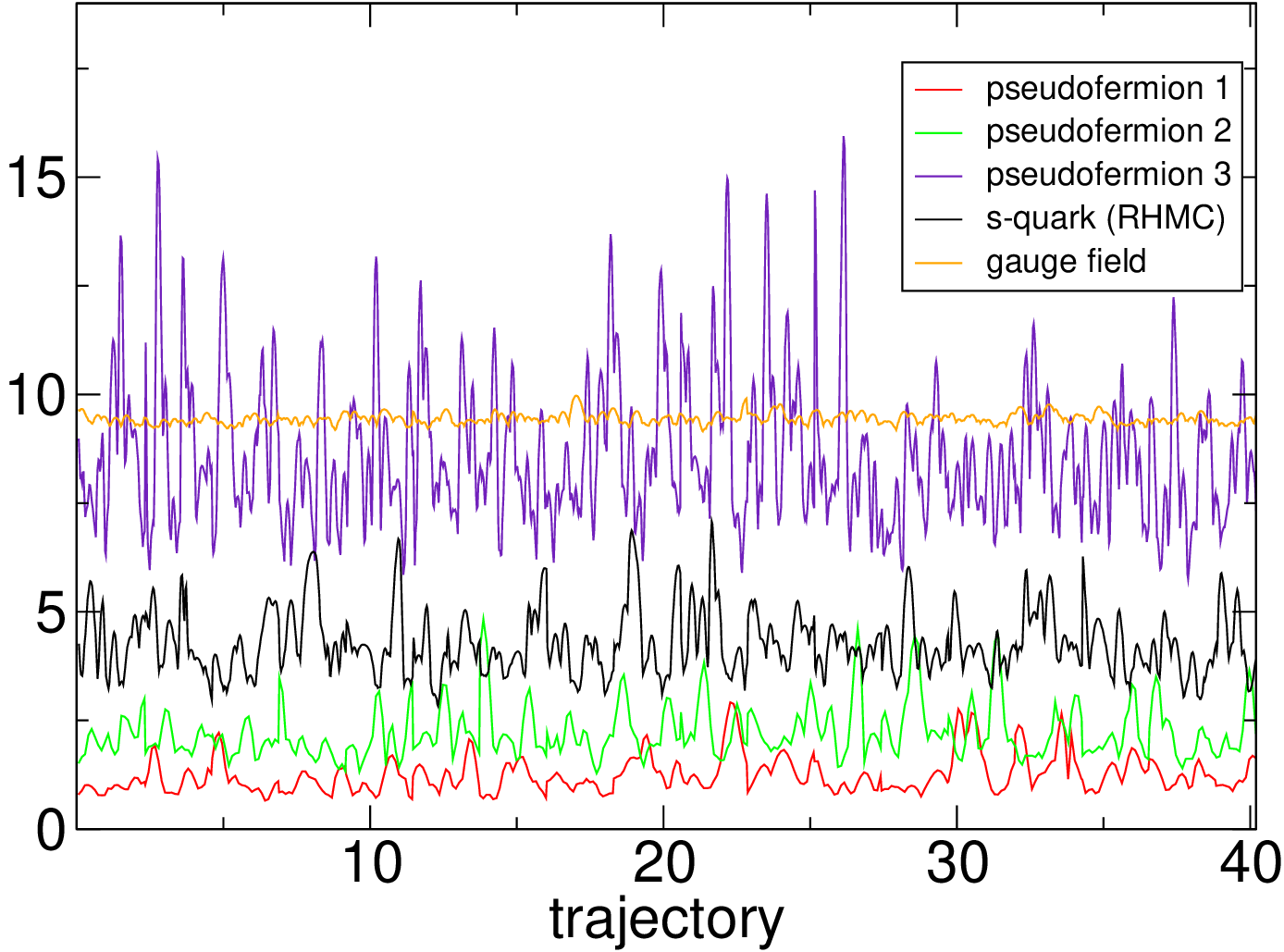}}
\caption{\label{ferm_force}
\dbs
Forces in the molecular dynamics time history.  We show here this
history for a typical sample of trajectories after thermalization.
Since the algorithm is more stable for large pion masses and spatial
sizes, we present --as a worst case scenario-- the fermionic force for
our smallest pion mass ($M_\pi{\approx}190$~MeV; $M_\pi L{\approx}4$).
The gauge force is the smoothest curve.  Then, from bottom to top
there are pseudofermion 1, 2, the strange quark and pseudofermion 3
forces, in order of decreasing mass.  No sign of instability is
observed.  }
\end{figure}

\begin{figure}
\centerline{\includegraphics[width=10cm]{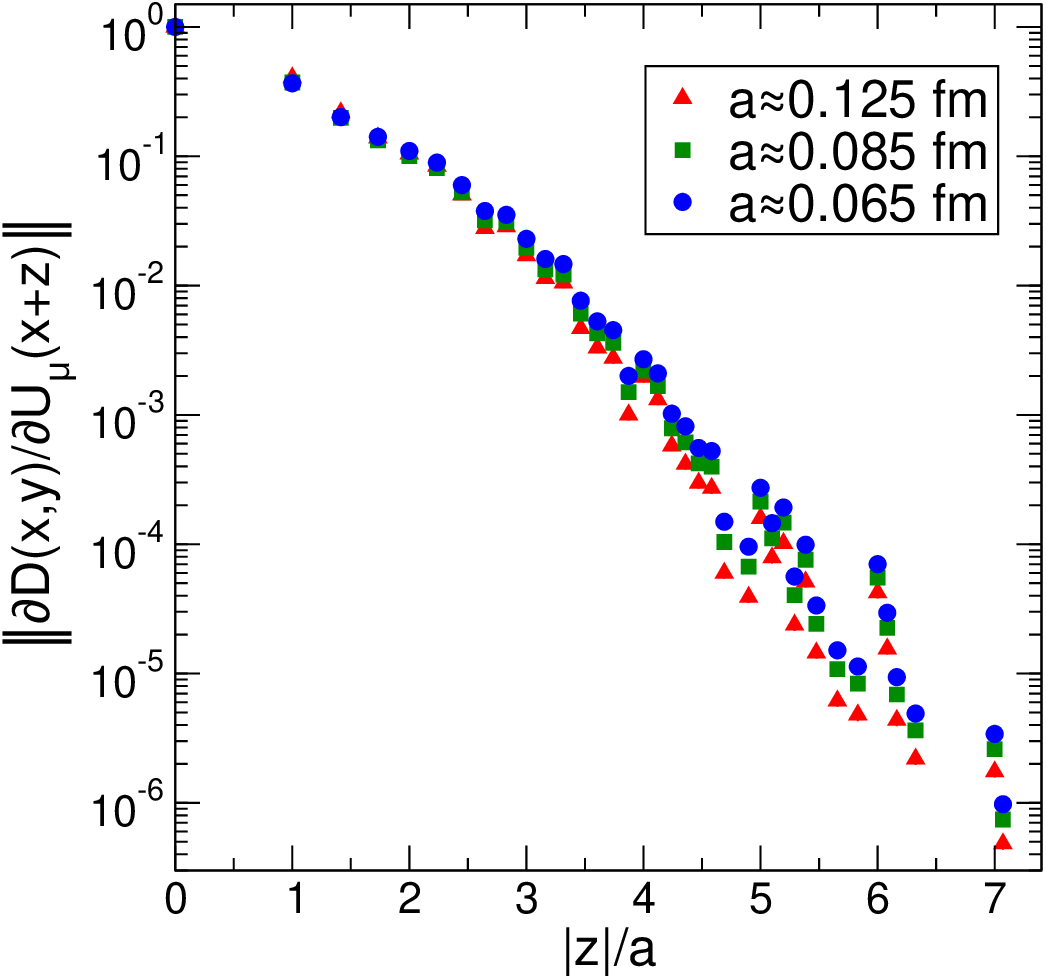}}
\caption{\label{locality} 
\dbs
Locality properties of the Dirac operator used in our simulations. 
In the literature, the term 
locality is used in two different ways (see e.g. 
\cite{S_Hernandez:1998et,S_Kovacs:2002nz,S_Durr:2005ax}). 
Our Dirac operator
is ultralocal in both senses. First of all (type A locality), in the sum 
$\sum_{x,y} {\bar \psi(x)} D(x,y)\psi(y)$
the non-diagonal elements of our $D(x,y)$ 
are by definition strictly zero for
all $(x,y)$ pairs except for nearest neighbors. The figure shows the second
aspect of locality (type B), i.e., 
how $D(x,y)$ depends on the gauge field $U_\mu$ at some
distance $z$: $\|\partial D(x,y)/\partial U_\mu(x+z)\|$. 
In the analyses we use the
Euclidian metric for $| z |$. We take the 
Frobenius norm of the resulting antihermitian matrix 
and sum over spin, color and Lorentz indices. 
An overall normalization is performed 
to ensure unity at $| z |$=0. 
The action is by definition ultralocal, thus
$\|\partial D(x,y)/\partial U_\mu(x+z)\|$
depends only on gauge field variables residing within a fixed range.
Furthermore, within this ultralocality range the decay is, in very good
approximation, exponential with an effective mass of about 2.2$a^{-1}$.
This is much larger than any of our masses, even on the coarsest lattices.
}
\end{figure}

The integrated autocorrelation times of the smeared
plaquette and that of the number of conjugate gradient iteration steps
are less than approximately ten trajectories. Thus every
tenth trajectory is used in the analysis.  We calculate the spectrum by using
up to eight timeslices as sources for the correlation functions. For
the precise form of the hadronic operators see
e.g. \cite{S_Montvay:1994cy}. We find that 
Gaussian sources and sinks of radii $\approx 0.32\,\fm$ 
are less
contaminated by excited states
than point sources/sinks (see Figure
\ref{sources}). The integrated autocorrelation times for hadron
propagators, computed on every tenth trajectory, are compatible with
0.5 and no further correlations were found through binning adjacent
configurations.  In order to exclude possible long-range correlations
in our simulations, we performed a run with 10,000 and one with 4,500
trajectories. No long-range correlations were observed. Further, we
never encountered algorithmic instabilities as illustrated by the time
history of the fermionic force in Figure \ref{ferm_force} and
discussed in more detail in \cite{S_Durr:2008rw}. Note that the
fermionic force, which is the derivative of the fermionic 
action with respect to
the gauge field, is directly related to the locality properties of our action
(see Figure
\ref{locality}).

\section*{Finite volume corrections and resonances}

For fixed bare parameters (gauge coupling, light quark mass and strange
quark mass), the energies of the different hadronic states depend on the
spatial size of the lattice (in a finite volume the energy spectrum is 
discrete  and all states are stable). There are two sources of volume
dependence, which we call type I and type II. These
were discussed in a series of papers by M. L\"uscher
\cite{S_Luscher:1985dn,S_Luscher:1986pf,S_Luscher:1990ux,S_Luscher:1991cf}. 
Both effects were quantified in a self-consistent manner 
in our analysis,
using only the results of our calculations
(i.e. no numerical inputs from experiments 
were used).

\begin{figure}
\centerline{
\includegraphics[width=8.15cm]{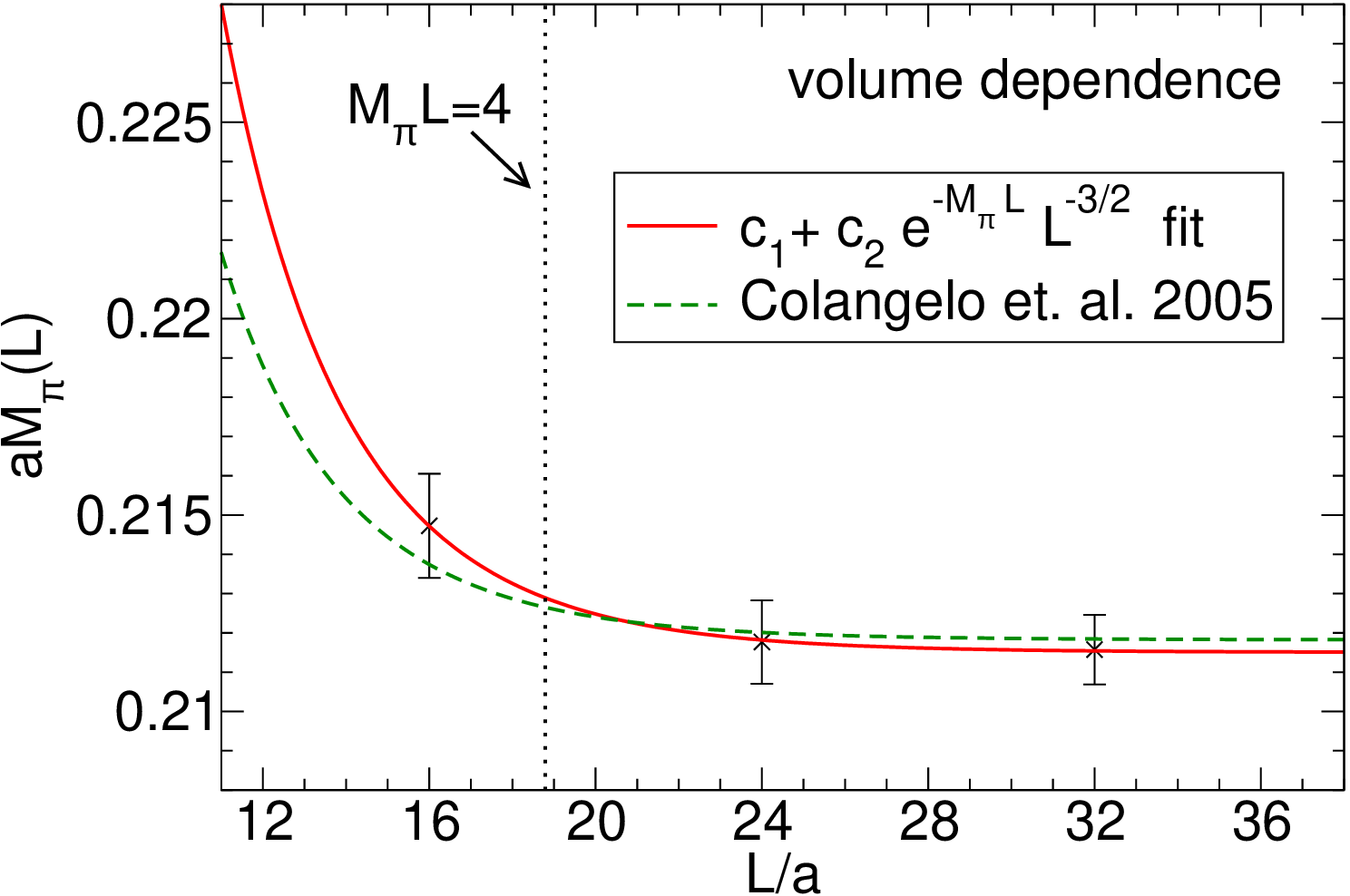}
\hspace*{0.1cm}
\includegraphics[width=8cm]{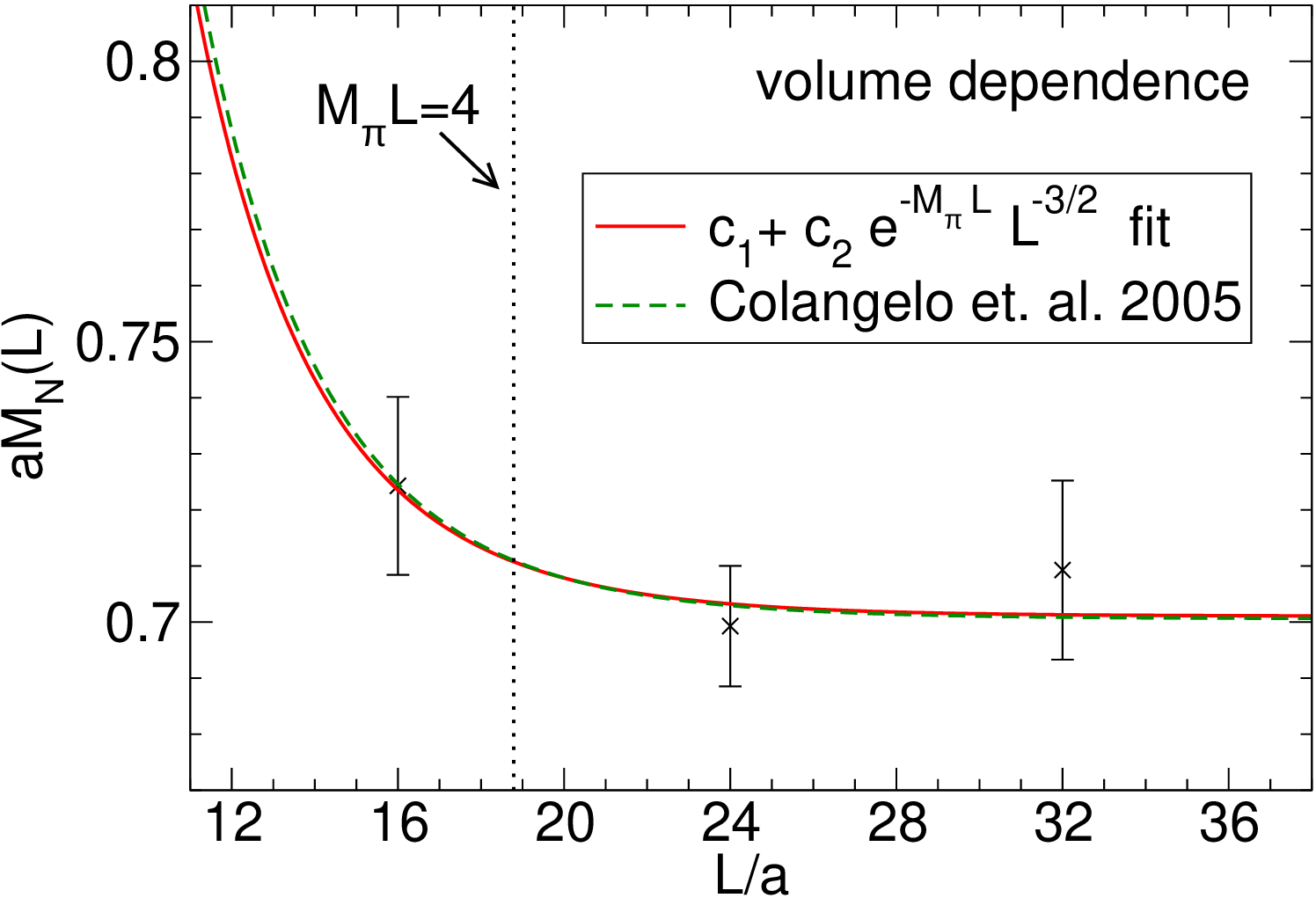}
}
\caption{\label{fin_volume}
\dbs
Volume dependence of the $\pi$ (left panel) and $N$ (right panel) masses 
for one of
our simulation points corresponding to $a\approx0.125 \fm$ and 
$M_\pi\approx 320$~MeV. The results of fits to the form 
$c_1+c_2 \exp(-M_\pi L)/(M_\pi L)^{3/2}$ are shown
as the solid curves, 
with $c_1=aM_X(L=\infty)$ and $c_2=ac_X(M_\pi)$ given in the text
($X=\pi, N$ for pion/nucleon).  The dashed curves correspond 
to fits with the $c_2$ of
refs.~\cite{S_Colangelo:2005gd,S_Colangelo:2005cg}.}
\end{figure}

Type I effects result from virtual pion exchanges between the
different copies of our periodic system. 
These effects induce
corrections in the spectrum which fall off exponentially with $M_\pi
L$ for large enough volumes \cite{S_Luscher:1985dn}. For one set of
parameters ($M_\pi{\approx}320$~MeV at
$a{\approx}0.125$~fm), additional 
runs have been carried out for several spatial
volumes ranging from $M_\pi L{\approx}3.5$ to 7. The size
dependences of the different hadron masses $M_X$ are successfully described
by
%
%
$M_X(L)=M_X+c_X(M_\pi)\cdot \exp(-M_\pi L)/(M_\pi L)^{3/2}$. 
Figure~\ref{fin_volume} shows the volume dependence 
at $M_\pi$=320~MeV 
for the two statistically most significant channels : the pion and
nucleon channels. The fitted $c_X$ coefficients are in good agreement
with those suggested by \cite{S_Colangelo:2005gd,S_Colangelo:2005cg}
which predicts a behavior of $c_X(M_\pi)\propto M_\pi^2$.  Our
results for these and other channels confirm the rule of thumb: $M_\pi
L{\gsim}4$ gives the infinite volume masses within statistical
accuracy. Nevertheless, we included these finite volume corrections in
our analysis.

The other source of volume dependence (type II) is relevant only to 
resonant states, in regions of parameter space where they would 
decay in infinite volume (five out of the twelve particles of the present
work are resonant states). 
Since in this case the lowest energy state with the quantum numbers
of the resonance in infinite volume is a two particle 
scattering state, we need to
take the effects of scattering states into account in our analysis.
For illustration we 
start by considering the hypothetical case where there is no
coupling between the resonance (which we will refer to as ``heavy
state'' in this paragraph) and the scattering states.
In a finite box of size $L$, the spectrum in the center of mass frame
consists of two particle states with energy $\sqrt{M_1^2+{\bf
    k}^2}+\sqrt{M_2^2+{\bf k}^2}$, where ${\bf k}={\bf n}2\pi/L$,
${\bf n}\in Z^3$ and $M_1$, $M_2$ are the masses of the lighter
particles (with corrections of type I discussed in the previous
paragraph) and, in addition, of the state of the heavy particle $M_X$
(again with type I corrections).  As we increase $L$, the energy of of
any one of the two particle states decreases and eventually becomes
smaller than the energy $M_X$ of $X$. An analogous phenomenon can
occur when we fix $L$ but reduce the quark mass (the energy of the two
light particles changes more than $M_X$).  In the presence of
interactions, this level crossing disappears and, due to the mixing of
the heavy state and the scattering state, an avoided level crossing
phenomenon is observed. Such mass shifts due to avoided level crossing
can distort the chiral extrapolation of hadron masses to the physical
pion mass.

The literature \cite{S_Luscher:1986pf,S_Luscher:1990ux,S_Luscher:1991cf}
provides a conceptually satisfactory basis to study resonances
in lattice QCD: each measured energy corresponds to a momentum, 
$|{\bf k}|$, 
which is a solution of a complicated non-linear equation. 
Though the necessary formulae can be found in the literature
(cf. equations (2.7, 2.10-2.13, 3.4, A3) of \cite{S_Luscher:1991cf}),
for completeness the main ingredients are summarized here. We follow
\cite{S_Luscher:1991cf} where the $\rho$-resonance was taken as an
example and it was pointed out that other resonances can be treated in
the same way without additional difficulties. The $\rho$-resonance
decays almost exclusively into two pions. The absolute value of the
pion momentum is denoted by $k=|{\bf k}|$.  The total energy of the
scattered particles is $W=2(M_\pi^2+k^2)^{1/2}$ in the center of mass
frame. The $\pi\pi$ scattering phase $\delta_{11}(k)$ in the isospin
$I=1$, spin $J=1$ channel passes through $\pi/2$ at the resonance
energy, which correspond to a pion momentum $k$ equal to
$k_\rho=(M_\rho^2/4-M_\pi^2)^{1/2}$. In the effective range formula
$(k^3/W)\cdot\cot \delta_{11}=a+bk^2$, this behavior implies
$a=-bk_\rho^2=4k_\rho^5/(M_\rho^2\Gamma_\rho)$, where $\Gamma_\rho$ is
the decay width the resonance (which can be parametrized by an
effective coupling between the pions and the $\rho$). The basic result
of \cite{S_Luscher:1990ux} is that the finite-volume energy spectrum
is still given by $W=2(M_\pi^2+k^2)^{1/2}$ but with $k$ being a
solution of a complicated non-linear equation, which involves the
$\pi\pi$ scattering phase $\delta_{11}(k)$ in the isospin $I=1$, spin
$J=1$ channel and reads $n\pi-\delta_{11}(k)=\phi(q)$. Here $k$ is in
the range $0<k<\sqrt{3}M_\pi$, $n$ is an integer, $q=kL/(2\pi)$ and
$\phi(q)$ is a known kinematical function which we evaluate
numerically for our analysis ($\phi(q)\propto q^3$ for small $q$ and
$\phi(q)\approx \pi q^2$ for $q\ge 0.1$ to a good approximation; more
details on $\phi(q)$ are given in Appendix A of
\cite{S_Luscher:1991cf}).  Solving the above equation leads to energy
levels for different volumes and pion masses (for plots of these
energy levels, see Figure 2 of \cite{S_Luscher:1991cf}).

Thus, the spectrum is determined by the box length $L$, the infinite
volume masses of the resonance $M_X$ and the two decay products $M_1$
and $M_2$ and one parameter, $g_X$, which describes the effective
coupling of the resonance to the two decay products and is thus
directly related to the width of the resonance.  In the unstable
channels our volumes and masses result in resonance states $M_X$ which
have lower energies than the scattering states (there are two
exceptions, see later). In these cases $M_X$ can be accurately
reconstructed from $L$, $M_1$, $M_2$ and $g_X$.
However, since we do not want to rely on experimental inputs in 
our calculations of the hadron masses, we choose to use, for each resonance,
our set of measurements for various $L$, $M_1$ and $M_2$ to determine
both $M_X$ and $g_X$.
With our choices of quark masses and volumes we find despite limited
sensitivity to the resonances' widths, that we can accurately determine the
resonances' masses. Moreover, the finite volume corrections
induced by these effects never exceed a few percent. In addition,
the widths obtained in the analysis are  
in agreement with the
experimental values, albeit with large errors. (For a precise determination
of the width, which is not our goal here, 
one would preferably need more than one energy level obtained
by cross-correlators. Such an analysis is 
beyond the scope of the present paper.)

Out of the 14$\cdot$12=168 
mass determinations (14 sets of lattice
parameters/volumes--see Table~\ref{parameters}--and 12 hadrons) 
there are two  cases for which
$M_X$ is larger than the energy of the lowest scattering state. 
These exceptions are
the $\rho$ and $\Delta$ for the lightest  pion mass  point at 
$a$$\approx$0.085~fm.
Calculating the energy levels according to 
\cite{S_Luscher:1990ux,S_Luscher:1991cf} for these two isolated cases, one
observes that the energy of the lowest lying state is already dominated by
the contribution from the neighboring, two particle state. More precisely,
this lowest state depends very weakly on the resonance mass, which
therefore cannot be extracted reliably. In fact, an extraction of $M_X$
from the lowest lying state would require precise information on the width
of the resonance.
%
Since one does not want to include the experimental width 
as an input in an ab initio calculation,
this point should not be used to determine $M_\rho$ and $M_\Delta$. 
Thus, for, and only for the $\rho$ and $\Delta$ channels, we left out this 
point from the analysis.

\section*{Approaching the physical mass point and the continuum limit}

We consider two different paths, in bare parameter space, to the
physical mass point and continuum limit. These correspond to two
different ways of normalizing the hadron masses obtained for a fixed
set of bare parameters.  For both methods we follow two strategies for
the extrapolation to the physical mass point and apply three different
cuts on the maximum pion mass. We also consider two different
parameterizations for the continuum extrapolation. All residual
extrapolation uncertainties are accounted for in the systematic
errors.  We carry out this analysis both for the $\Xi$ and for
the $\Omega$ sets separately. 

We call the two ways
of normalizing the hadron masses: 1. ``the ratio method'',
2. ``mass independent scale setting''. 

1. The ratio method is motivated by the fact that 
in QCD one can calculate only dimensionless combinations of
observables, e.g. mass ratios. Furthermore, in such ratios cancellations
of statistitical uncertainties and systematic effects may occur.
The method uses
the ratios 
$r_X$=$M_X$/$M_\Xi$ and parametrizes the mass dependence of these
ratios in terms of $r_\pi$=$M_\pi$/$M_\Xi$ and $r_K$=$M_K$/$M_\Xi$. 
The continuum extrapolated two-dimensional surface
$r_X$=$r_X$($r_\pi$,$r_K$) is an unambiguous prediction of QCD for a
particle of type $X$ (a couple of points of this surface have been
determined in \cite{S_Durr:2008rw}). One-dimensional slices
($2r_K^2-r_\pi^2$ was set to 0.27, to its physical value) of the
two-dimensional surfaces for $N$ and $\Omega$ are shown on
Figure 2 of our paper. 
(Here we write the formulas relevant for $\Xi$ set;
analogous expressions hold for the $\Omega$ set. The final
results are also given for the $\Omega$ set).

A linear term in $r_K^2$ (or $M_K^2$) is sufficient for the small
interpolation needed in the strange quark mass direction.
On the other hand, our data is accurate enough that some curvature with 
respect to $r_\pi^2$ (or $M_\pi^2$)
is visible in some channels. In order to perform an extrapolation to 
the physical pion mass one needs to use an expansion  
around some pion mass point. 
This point can be $r_\pi{=}0$ ($M_\pi{=}0$), 
which corresponds to 
chiral perturbation theory. Alternatively one can use a non-singular point
which is in a range of
$r_\pi^2$ (or $M_\pi^2$) which includes the physical and simulated pion masses.
We follow both strategies
(we call them ``chiral fit'' and ``Taylor fit'', respectively). 

In addition to a linear expression in $M_\pi^2$, chiral perturbation theory
predicts~\cite{S_Langacker:1974bc} an $M_\pi^3$ next-to-leading
order behavior for masses other than those of the pseudo-Goldstone bosons.
This provides our first strategy (``chiral fit'').  A generic
expansion of the ratio $r_X$ around a reference point reads:
$r_X=r_X(ref)+\alpha_X[r_\pi^2-r_\pi^2(ref)]+\beta_X[r_K^2-r_K^2(ref)]+hoc$,
where $hoc$ denotes higher order contributions. In our chiral fit,
$hoc$ is of the form $r_\pi^3$, all coefficients are left free and the
reference point is taken to be $r_\pi^2(ref){=}0$ and $r_K^2(ref)$ is
the midpoint between our two values of $r_K^2$, which straddle
$r_K^2(phys)$.  The
second strategy is a Taylor expansion in $r_\pi^2$ and $r_K^2$ around
a reference point which does not correspond to any sort of singularity
(``Taylor fit'').
In this case, $r_K^2(ref)$ is again at the center of our fit range and
$r_\pi^2(ref)$ is the midpoint of region defined by the physical value
of the pion mass and the largest simulated pion mass considered.  This
choice guarantees that all our points are well within the radius of
convergence of the expansion, since the nearest singularities are at
$M_\pi=0$ and/or $M_K=0$.  Higher order contributions, $hoc$, of the
form $r_\pi^4$ turned out to be sufficient.
%

We extrapolate to the physical pion mass following both strategies
(cubic term of the ``chiral fit'' or a quartic contribution of the
``Taylor fit''). The variations in our results which follow from the
use of these different procedures are included in our systematic error
analysis.

The range of applicability of these expansions is not precisely known a priori. 
In case of the two vector mesons the coefficients of the
higher order ($r_\pi^3$ or $r_\pi^4$)
contributions were consistent with zero even when using our
full pion mass range. 
Nevertheless, they are included in the analysis. 
For the baryons, however, the higher order contributions are
significant. The difference between the results obtained with the two
approaches gives some indication of the possible contributions of yet
higher order terms not included in our fits. To quantify these
contributions further, we consider three different ranges of pion
mass.  In the first one we include all 14 simulation points, in the
second one we keep points upto $r_\pi=0.38$ (thus dropping two pion
mass points) and in the third one we apply an even stricter cut at
$r_\pi=0.31$ (which corresponds to omitting the five heaviest points).
The pion masses which correspond to these cuts will be given
shortly. The differences between results obtained using these three
pion mass ranges are included in the systematic error analysis.

To summarize, the ``ratio method'' 
uses the input data $r_X$, $r_\pi$ and $r_K$ to determine
$r_X(ref)$, $\alpha_X$ and $\beta_X$ and, based on them, we obtain
$r_X$ at the physical point. The determination of this value is done
with the two fit strategies (``chiral'' and ``Taylor'')  
for all three pion mass ranges.

2. The second, more conventional method (``mass independent scale setting'') 
consists of first setting the 
lattice spacing by
extrapolating $M_\Xi$ to the physical point, given by the physical
ratios of $M_\pi/M_\Xi$ and $M_K/M_\Xi$. 
Using the resulting lattice spacings
obtained for each bare gauge coupling,
we then proceed to fit $M_X$ vs. $M_\pi$ and $M_K$ applying both
extrapolation stratagies (``chiral'' and ``Taylor'') discussed above. 
We use the same three pion mass ranges as for the ``ratio method'':
in the first all simulation points are kept, in the second we cut
at $M_\pi{=}560$~MeV and the third case this cut was brought down
to $M_\pi{=}450$~MeV. 

\bigskip

As shown in the $2{+}1$ flavor scaling study of \cite{S_Durr:2008rw},
typical hadron masses, obtained in calculations which are performed
with our $O(a)$-improved action, deviate from their continuum values
by less than approximately 1\% for lattice spacings up to $a\approx
0.125\,\mathrm{fm}$. Moreover, \cite{S_Durr:2008rw} shows that these
cutoff effects are linear in $a^2$ as $a^2$ is scaled from $a\sim
0.065\,\mathrm{fm}$ to $a\sim 0.125\,\mathrm{fm}$ and even
above. Thus, we use the results obtained here, for three values of the
lattice spacing down to $a\sim 0.065\,\mathrm{fm}$, to extrapolate
away these small cutoff effects, by allowing $r_X(ref)$ (or
$M_X(ref)$) to acquire a linear dependence in $a^2$.  In addition to
the extrapolation in $a^2$, we perform an extrapolation in $a$ and use
the difference as an estimate for possible contributions of higher
order terms not accounted for in our continuum extrapolation.

The physical mass and continuum extrapolations are carried out
simultaneously in a combined, correlated analysis.

\section*{Statistical and systematic error analysis}


Systematic uncertainties are accounted for as described above. In addition,
to estimate the possible contributions of excited states to our extraction of
hadron masses from the time-dependence of two-point functions, we
consider 18 possible time intervals whose initial time varies from low
values, where excited states may contribute, to higher values, where
the quality of fit clearly indicate the absence of such contributions.

Since the light hadron spectrum is known experimentally it is of
extreme importance to carry out a blind data analysis. One should
avoid any arbitrariness related e.g. to the choice of some fitting
intervals or pre-specified coefficients of the chiral fit. We follow
an extended frequentist's method \cite{S_Yao:2006px}. To this end we
combine several possible sets of fitting procedures (without imposing
any additional information for the fits) and weight them according to
their fit quality.  Thus, we have 2 normalization methods, 2
strategies to extrapolate to the physical pion mass, 3 pion mass
ranges, 2 different continuum extrapolations and 18 time intervals for
the fits of two point functions, which result in
2$\cdot$2$\cdot$3$\cdot$2$\cdot$18=432 different results for the mass
of each hadron.

In lattice QCD calculations, electromagnetic interactions are absent
and isospin is an exact symmetry. Electromagnetic and isospin breaking
effects are small, typically a fraction of 1\% in the masses of light
vector mesons and baryons~\cite{S_Gasser:1982ap}. Moreover,
electromagnetic effects are a small fraction of the mass difference
between the members of a same isospin
multiplet~\cite{S_Gasser:1982ap}. We account for these effects by
isospin averaging the experimental masses to which we compare our
results. This eliminates the leading isospin breaking term, leaving
behind effects which are only a small fraction of 1\%. For the pion
and kaon masses, we use isospin averaging and Dashen's
theorem~\cite{S_Dashen:1969eg}, which determines the leading order
electromagnetic contributions to these masses. Higher order
corrections, which we neglect in our work, are expected to be below
the 3 per mil level (see e.g. \cite{S_Aubin:2004fs}). All of these
residual effects are very small, and it is safe to neglect them in
comparing our results to experiment.

\begin{figure}
\centerline{
\includegraphics[width=12cm]{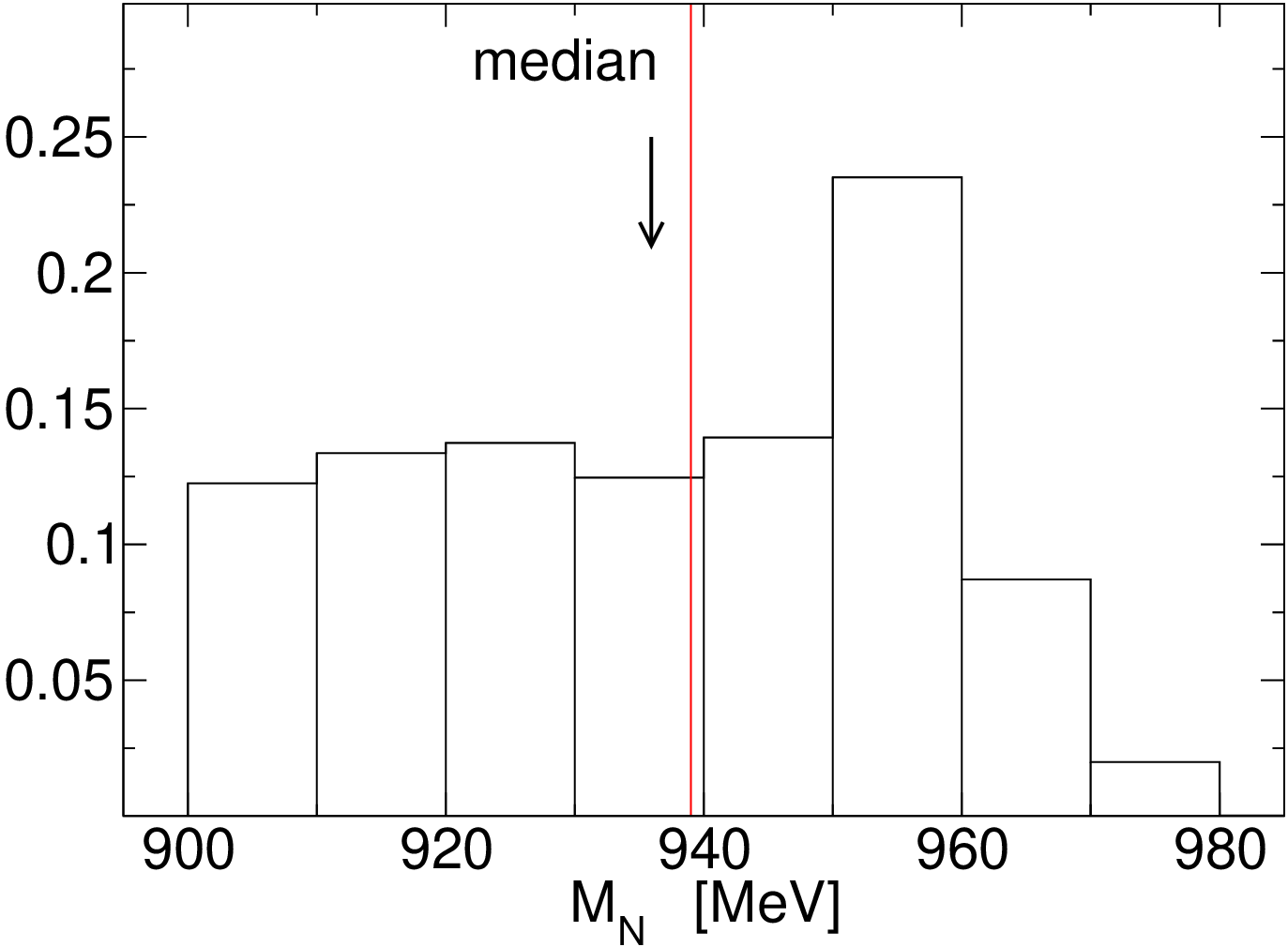}
}
\caption{\label{err_distr} 
\dbs
Distribution used to estimate the central value and systematic error on the 
nucleon mass. The distribution was
obtained from 432 different fitting procedures as
explained in the text.
The median is shown by 
the arrow. The experimental value of the nucleon mass is indicated by the 
vertical 
line.
}
\end{figure}

The central value and systematic error bar for each hadron mass is
determined from the distribution of the results obtained from our 432
procedures, each weighted by the corresponding fit quality. This
distribution for the nucleon is shown in Figure \ref{err_distr}. The
central value for each hadron mass is chosen to be the median of the
corresponding distribution. The systematic error is obtained from the
central 68\% confidence interval. To calculate statistical errors, we
repeat the construction of these distributions for 2000 bootstrap
samples. We then build the bootstrap distribution of the medians of
these 2000 distributions.  The statistical error (SEM) on a hadron
mass is given by the central 68\% confidence interval of the
corresponding bootstrap distribution. These systematic and statistical
errors are added in quadrature, yielding our final error bars. The
individual components of the total systematic error are given in
Table~\ref{error_budget}.



\clearpage

\begin{table} 
\begin{center}
\begin{tabular}{|l|l|l|l|l|}
\hline
$\beta$& $am_{ud}$&  $am_s$ & $L^3\cdot T$ & \# 
traj.\\
  \hline\hline
  \multirow{5}{*}{3.3}
  & -0.0960 & -0.057 & $16^3\cdot 32$ & 10000 \\
  & -0.1100 & -0.057 & $16^3\cdot 32$ & 1450\\
  & -0.1200 & -0.057 & $16^3\cdot 64$ & 4500 \\
  & -0.1233 & -0.057 & $16^3\cdot 64$ / $24^3\cdot 64$ / $32^3\cdot 64$
                                & 5000 / 2000 / 1300 \\
  & -0.1265 & -0.057 & $24^3\cdot 64$ & 2100 \\
  \hline
  \multirow{4}{*}{3.57} 
  & -0.0318 & 0.0 / -0.01 &  $24^3\cdot 64$ & 1650 / 1650\\
  & -0.0380 & 0.0 / -0.01 & $24^3\cdot 64$  & 1350 / 1550\\
  & -0.0440 & 0.0 / -0.007& $32^3\cdot 64$  & 1000 / 1000\\
  & -0.0483 & 0.0 / -0.007& $48^3\cdot 64$  & 500 / 1000\\
  \hline
  \multirow{5}{*}{3.7} 
  & -0.0070 & 0.0 & $32^3\cdot 96$ & 1100\\
  & -0.0130 & 0.0 & $32^3\cdot 96$ & 1450\\
  & -0.0200 & 0.0 & $32^3\cdot 96$ & 2050\\
  & -0.0220 & 0.0 & $32^3\cdot 96$ & 1350\\
  & -0.0250 & 0.0 & $40^3\cdot 96$ & 1450\\
\hline
\end{tabular}
\end{center}
\caption{\label{parameters}  
\dbs
Bare lagrangian parameters, lattice sizes and statistics.
The table summarizes the 14 simulation points at three different 
lattice spacings ordered by the light quark masses.
Note that due to the additive mass renormalization, the bare mass 
parameters can be negative.
At each lattice spacing 4-5 light quark masses are studied. 
The results of all these 
simulations are used to perform a combined mass and continuum extrapolation to 
the physical point.
In addition,
for one set of Lagrangian parameters, different volumes were studied and 
four of our simulations at $\beta$=3.57 were repeated with different 
strange quark masses. 
}
\end{table}

\begin{table} 
\begin{center}
\begin{tabular}{|l|l|l|l|l|}
\hline
& continuum extrapolation & chiral fits/normalization & excited states& finite volume \\
  \hline\hline
$\rho$ & 0.20 & 0.55 & 0.45 & 0.20 \\
$K^*$ & 0.40 & 0.30 & 0.65 & 0.20 \\
$N$ & 0.15 & 0.90 & 0.25 & 0.05 \\
$\Lambda$ & 0.55 & 0.60 & 0.40 & 0.10 \\
$\Sigma$ & 0.15 & 0.85 & 0.25 & 0.05 \\
$\Xi$ & 0.60 & 0.40 & 0.60 & 0.10 \\
$\Delta$ & 0.35 & 0.65 & 0.95 & 0.05 \\
$\Sigma^*$ & 0.20 & 0.65 & 0.75 & 0.10 \\
$\Xi^*$ & 0.35 & 0.75 & 0.75 & 0.30 \\
$\Omega$ & 0.45 & 0.55 & 0.60 & 0.05 \\
\hline
\end{tabular}
\end{center}
\caption{\label{error_budget}\dbs
Error budget given as fractions of the total systematic error. Results
represent averages over the $\Xi$ and $\Omega$ sets.  The columns
correspond to the uncertainties related to the continuum extrapolation
(${\cal O}(a)$ or ${\cal O}(a^2)$ behavior), to the extrapolation to
the physical pion mass (obtained from chiral/Taylor extrapolations for
each of three possible pion mass intervals using the ratio method or
the mass independent scale setting), to possible excited state
contamination (obtained from different fit ranges in the mass
extractions), and to finite volume corrections (obtained by including
or not including the leading exponential correction). If combined in
quadrature, the individual fractions do not add up to exactly 1.  The
small ($\lsim 20\%$) differences are due to correlations, the
non-Gaussian nature of the distributions and the fact that the very
small finite volume effects are treated like corrections in our
analysis, not contributions to the systematic error (the effect of yet
higher order corrections is completely negligible). The finite volume
corrections of the decuplet resonances increase with increasing
strange content. This is only due to the fact that these are fractions
of decreasing total systematic errors.  The absolute finite volume
corrections of these resonances are on the same level.
}
\end{table}

\clearpage

\makeatletter \renewcommand\@biblabel[1]{S#1.} \makeatother

\end{document}